\documentclass{aa}

\usepackage{graphics}
\begin{document}

   \thesaurus{06         
              (02.01.2;  
	       08.02.2;  
               08.02.4;  
	       08.09.2 WX Ari;  
               08.14.2)  
	       }  
  \title{Long-term photometry of WX Arietis: evidence for eclipses and dips\thanks{Based on 
observations made with the Optical Ground Station (OGS) and the IAC--80 
telescopes, operated on the island of Tenerife by the European Space Agency 
(ESA) and the Ins\-ti\-tu\-to de Astrof\'\i sica de Canarias (IAC), 
respectively, at the Spanish Observatorio del Teide 
and with the Jacobus Kapteyn Telescope (JKT), operated on the island of La Palma 
by PPARC at the Spanish Observatorio del Roque de los Muchachos.}}

  \author{P. Rodr\'\i guez--Gil
          \inst{1}
          \and
           J. Casares\inst{1}
	  \and
	  V. S. Dhillon\inst{2}
          \and
	  I. G. Mart\'\i nez--Pais\inst{1}
          }

   \offprints{P. Rodr\'\i guez--Gil
   \\ e-mail: prguez@ll.iac.es}

   \institute{Instituto de Astrof\'\i sica de Canarias, V\'\i a L\'actea s/n, La 
Laguna, 38200, Santa Cruz de Tenerife, Spain
         \and
 Department of Physics and Astronomy, University of Sheffield, Sheffield S3 7RH, 
UK
             }

   \date{Received/ Accepted 8 November 1999}

   \titlerunning{Photometry of WX Arietis: eclipses and dips}

   \maketitle

   \begin{abstract}
   
We present R-band photometry of the \object{SW Sex}-type cataclysmic variable
WX Arietis made in October 1995 and August 1998--February 1999. Contrary to 
previous results, we find that \object{WX Ari} is an eclipsing system with an 
orbital inclination of $i\simeq72^{\rm o}$. The R-band light curves display highly 
variable, shallow eclipses $\sim$0.15-mag deep and $\simeq$40 min long. The
observed eclipse depth suggests a partial eclipse of the accretion disc. The light 
curves also show a wide dip in brightness centred at orbital phase 
$\varphi\sim0.75$ and a hump close to the opposite phase at $\varphi\sim0.2$. 
The observed dip may be explained by the probable vertical thickening 
of the outer rim of the accretion disc downstream from the bright spot.
We also demonstrate that the disc brightness in all \object{SW Sex} systems is nearly the same. This implies that
the orbital inclination of these systems is only a function of eclipse depth. 

   \end{abstract}


\section{Introduction}

The \object{SW Sex} phenomenon, a term coined by Thorstensen et al. 
(\cite{thorstensen}), is common to a set of nova-like cataclysmic variables 
(CVs) which share the same peculiar but consistent spectroscopic behaviour. This 
group of CVs originally comprised only eclipsing systems with orbital periods in
the range 3--4 hr. All of these systems exhibit strong single-peaked Balmer,
\ion{He}{i} and \ion{He}{ii} emission lines which, with the exception of
\ion{He}{ii}, remain largely unobscured during primary eclipse. In addition, the 
radial velocity curves of Balmer and \ion{He}{i} lines show significant phase 
lags relative to the photometric ephemeris. Furthermore, these emission lines 
display absorption components which show maximum depth at phases
opposite primary eclipse (Szkody \& Pich\'e \cite{szkody90}).  

Given that all of the original members of the \object{SW Sex} class are 
eclipsing CVs, a number of authors have speculated that the various peculiar 
phenomena exhibited by the \object{SW Sex} stars are a consequence of their high 
orbital inclinations. This assumption was thrown into 
doubt when a number of supposed low-inclination members of the class were 
discovered, including \object{WX Ari} (Beuermann et al. \cite{beuermann92}; 
Hellier et al. \cite{hellier94b}), \object{V795 Her} (Casares et al. 
\cite{casares96}; Dickinson et al. \cite{dickinson97}), \object{S193} (Rodr\'\i 
guez-Gil et al. \cite{rodriguez98}; Mart\'\i nez-Pais et al. \cite{martinez99};
Taylor et al. \cite{taylor99}) 
and \object{V442 Oph} (Hoard \& Szkody \cite{hoard99}). Although these systems 
present the same spectroscopic properties as their high-inclination partners 
(with the exception of the occurrence of eclipses), there is still some 
controversy about whether they really are the low-inclination counterparts of 
\object{SW Sex} stars. In this paper we will show that the best example of a 
proposed low-inclination \object{SW Sex} star, \object{WX Ari}, exhibits 
eclipses and is actually a high-inclination \object{SW Sex} star.

\object{WX Ari} (also known as \object{PG 0244+104}) is one of a number of CVs 
discovered by the Palomar-Green (PG) survey for ultraviolet excess objects 
(Green et al. \cite{green82}). Beuermann et al. (\cite{beuermann92}) undertook 
the first spectroscopic study of this star, finding an orbital period of 
0.13934$\pm$0.00006 d. This work also showed that \object{WX Ari} displays many 
of the characteristic peculiarities of the \object{SW Sex} systems. After 
detecting the absence of eclipses, the authors suggested that \object{WX Ari} 
could be a low-inclination \object{SW Sex} star. Until then, the only 
photometric data available were those of Warner (\cite{warner83}), which did not 
cover the 3.34 hr orbital period of the system, so Hellier et al. 
(\cite{hellier94b}) decided to obtain new photometry to settle the issue of
whether \object{WX Ari} eclipses. They concluded that \object{WX Ari} is a low-
inclination \object{SW Sex} star not showing eclipses. Curiously, the V-band 
light curve presented in the upper panel of their Fig. 1 {\it does} show a 
shallow eclipse. This led Dhillon (\cite{dhillon96}) to propose that \object{WX 
Ari} is not a low-inclination system. Since Hellier's database only spanned two 
observing nights, we decided to undertake a new photometric study of \object{WX 
Ari} with the aim of giving a definite answer to the eclipsing nature of this 
\object{SW Sex} star. The results of this study are presented in this paper.


\section{Observations and data reduction}

The photometric data were obtained with the 1-m Optical Ground Station (OGS) and 
the 0.82-m IAC--80 telescopes at the Observatorio del Teide on Tenerife, and 
with the 1-m Jacobus Kapteyn Telescope (JKT) at the Observatorio del Roque de 
los Muchachos on La Palma. The system was observed on one night in October 1995 
with the JKT and for a total of 17 nights in the period August 1998--February 
1999 with the OGS and the IAC--80 telescopes. The observations on both 
telescopes on Tenerife were made using CCD cameras equipped with Thomson 
1024$\times$1024-pixel$^2$ chips, whilst the JKT data were obtained with a Tek 
1024$\times$1024-pixel$^2$ chip. A detailed observing log is given in 
Table~\ref{table1}. All the data were acquired through a Kron-Cousins R-filter 
and the time resolution was always better than 90 s.

   \begin{table}
      \caption[]{Observing log}
         \label{table1}
      \[
         \begin{array}{ccccc}
            \hline
            \noalign{\smallskip}
            Night      &  Telescope  &  Filter  & Exposure~(s)  & Coverage~(h)  
\\
            \noalign{\smallskip}
            \hline
            \noalign{\smallskip}
            1995~Oct~15 & JKT & R & 30 & 4.91     \\
            1998~Aug~26 & OGS & R & 60 & 1.17     \\
            1998~Aug~27 & OGS & R & 60 & 2.51     \\
	    1998~Sep~02 & IAC 80 & R & 40 & 3.46   \\
	    1998~Sep~03 & IAC 80 & R & 50 & 4.78   \\
	    1998~Sep~23 & OGS & R & 40 & 2.77     \\
	    1998~Sep~24 & OGS & R & 25 & 6.88     \\
	    1998~Oct~05 & IAC 80 & R & 60 & 3.39   \\
            1998~Nov~24 & IAC 80 & R & 40 & 3.60   \\
            1998~Dec~09 & IAC 80 & R & 40 & 3.88   \\
            1999~Jan~30 & IAC 80 & R & 40 & 3.70   \\
            1999~Jan~31 & IAC 80 & R & 40 & 3.60   \\
            1999~Feb~01 & IAC 80 & R & 50 & 3.42   \\
            1999~Feb~02 & IAC 80 & R & 50 & 3.64   \\
            1999~Feb~05 & IAC 80 & R & 70 & 3.42   \\
            1999~Feb~06 & IAC 80 & R & 40 & 3.45   \\
            1999~Feb~07 & IAC 80 & R & 40 & 1.51   \\
            1999~Feb~09 & IAC 80 & R & 50 & 3.06   \\
            
            \noalign{\smallskip}
            \hline
         \end{array}
      \]
   \end{table}

The individual images were bias-corrected and then flat-fielded in the standard 
way. All the data reduction was undertaken with the {\sc IRAF}\footnote{{\sc 
iraf} is distributed by the National Optical Astronomy Observatories, which is 
operated by the Association of Universities for Research in Astronomy, Inc., 
under contract with the National Science Foundation.} package. The instrumental
magnitudes were then found by performing aperture photometry on the individual 
images. The seeing during\par

\begin{figure}[!ht]
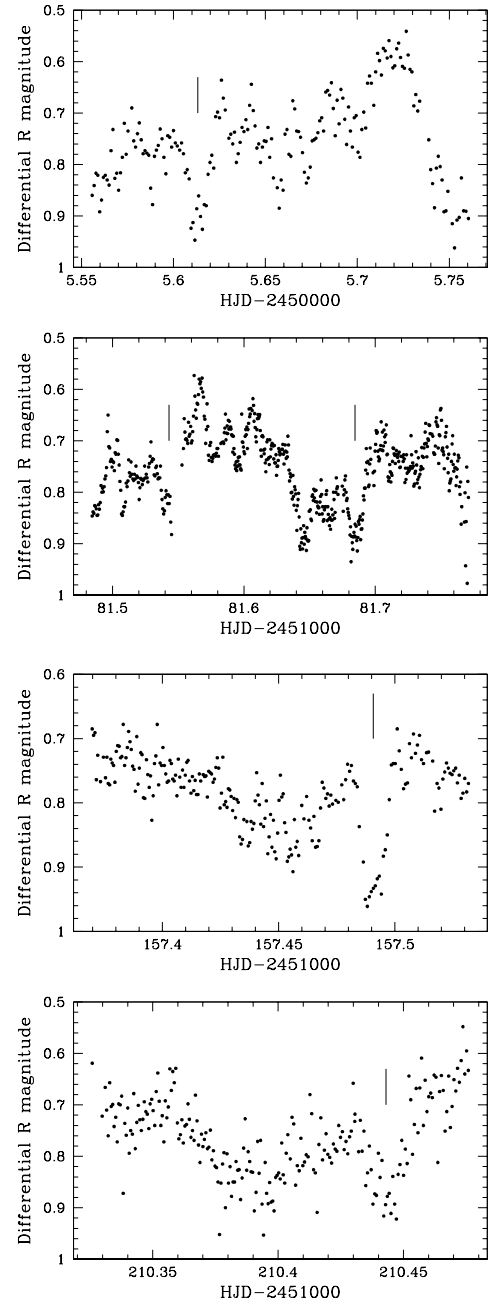

  \vspace{0.2cm}
  \resizebox{\hsize}{!}{\includegraphics{h1808.f1a}}
  \resizebox{\hsize}{!}{\includegraphics{h1808.f1b}}
   \caption{Selected light curves of \object{WX Ari} in the R-band. Five 
eclipses are marked. Although flickering is very strong the eclipses are easily 
seen.}
   \label{fig1}
\end{figure}

\noindent
all the observing runs was estimated to be always 
better than 1.5\arcsec, even reaching sub-arcsecond values on a couple of 
nights. From the scatter in the comparison star light curves we estimate that 
the differential photometry is accurate to $\sim$1 per cent.


\section{Results and discussion}

\subsection{Evidence of eclipses and ephemeris}

In Fig.~\ref{fig1} we present four R-band light curves of \object{WX Ari}. From 
the light curves it is clear that \object{WX Ari} displays eclipses 
approximately 0.15-mag deep, in contrast with the previous results of Beuermann 
et al. (\cite{beuermann92}) and Hellier et al. (\cite{hellier94b}). Due to the shallowness of the observed
eclipses, we consider that they are partial eclipses of the accretion disc (see
below). The duration 
of the eclipses is $\simeq$40 min.

We derived the first photometric ephemeris for \object{WX Ari} from 13 eclipses. The times of mid-eclipse were calculated 
by fitting parabolas to the bottom of the eclipses. A least squares fit to these 
eclipse timings yields the following ephemeris:

\begin{equation}
\label{eq1}
T_{0}={\rm HJD}~2451081.54406(2)~+~0.13935119(3)~E
\end{equation}

The eclipse timings and the differences between the observed and calculated 
times of mid-eclipse (O-C) are given in Table~\ref{table2}. There is no evidence 
for any periodicity in the O-C diagram.

   \begin{table}
      \caption[]{Times of mid-eclipse for \object{WX Ari}}
         \label{table2}
      \[
         \begin{array}{ccc}
            \hline
            \noalign{\smallskip}
            \rm HJD~(mid-eclipse)      &  \rm{Cycle~Number} &  $O-C$  \\
            \noalign{\smallskip}
            (2450000$+$)            &     (E)       &   (s)   \\
            \noalign{\smallskip}
            \hline
            \noalign{\smallskip}
             5.61319    & -7721 & -28.37  \\
             1053.67288 & -200  & -81.39  \\
             1059.66493 & -157  & -85.81 \\
             1060.63953 & -150  & -159.97 \\
             1080.70952 & -6    & 135.40 \\
             1081.54740 & 0     & 288.58 \\
             1081.68478 & 1     & 118.27 \\
             1142.44301 & 437   & 214.27 \\
             1142.57998 & 438   & 8.85 \\
             1157.49072 & 545   & 22.59 \\
             1209.46623 & 918   & -192.02 \\
             1210.43840 & 925   & -476.13 \\
             1215.46330 & 961   & 237.29 \\
              \noalign{\smallskip}
             \hline 
         \end{array}
      \]
   \end{table}

\subsection{Light curve morphology and eclipse profiles}

The individual light curves are dominated by intense flickering, typically 
associated with emission from the bright spot and/or turbulent regions of the 
disc (see e.g. Horne \& Stiening \cite{horne85}).

\object{WX Ari} displays asymmetric eclipses with ingress steeper than egress. 
This asymmetry in eclipse shape is a common feature to all \object{SW Sex} 
systems and is indicative of the bright spot (Penning et al. \cite{penning84}). Only \object{PX And} exhibits a 
different behaviour in its continuum light curves, with eclipse egress steeper 
than ingress (Thorstensen et al. \cite{thorstensen}). Eclipses in \object{WX 
Ari} are shallower ($\sim$0.15 mag) than in the lowest-inclination eclipsing 
\object{SW Sex} star, \object{PX And} ($\sim$0.5 mag, and variable). The 
eclipses are also variable in depth, a feature also observed in \object{PX And} 
(Thorstensen et al. \cite{thorstensen}). The variable depth of the eclipse in 
CVs is still a puzzling matter. This variability may be related to local 
fluctuations in the disc --which is in 
a time-averaged steady state-- caused for instance by a rotating accretion 
curtain (Thorstensen et al. \cite{thorstensen}). 

\begin{figure}
  \resizebox{\hsize}{!}{\includegraphics{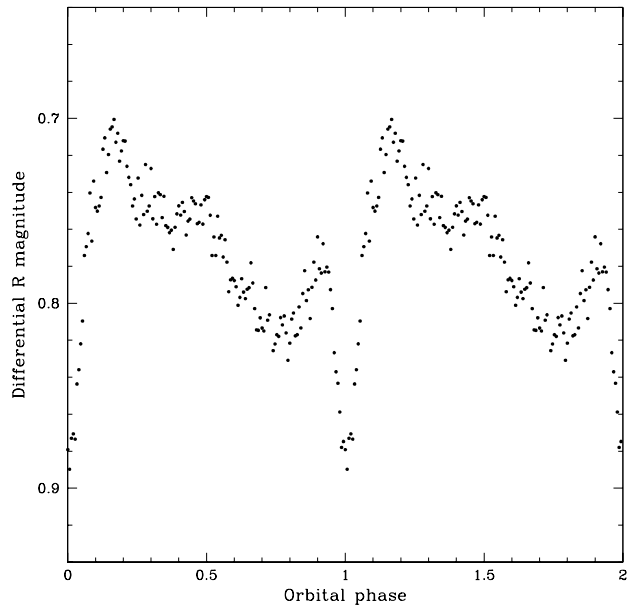}}
   \caption{The mean R-band light curve of \object{WX Ari} after binning the 
data into 150 phase bins. Note the well-defined eclipse at zero phase, the hump 
at phase $\sim$0.2 and the dip at phase $\sim$0.75. Two orbital periods are 
plotted for clarity.}
   \label{fig2}
\end{figure}

\begin{figure}
  \resizebox{\hsize}{!}{\includegraphics{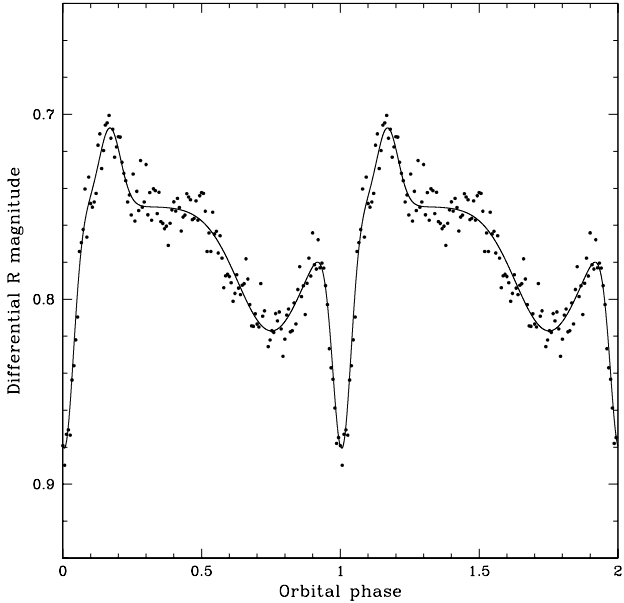}}
   \caption{Triple Gaussian fit to the mean R-band light curve of \object{WX 
Ari} ({\em solid line}).}
   \label{fig3}
\end{figure}

The mean R-band light curve of \object{WX Ari} is presented in Fig.~\ref{fig2} 
after phase-binning the data into 150 bins. The light curve shows an
intricate behaviour: apart from the well-defined eclipse, it exhibits 
a large dip in brightness centred around phase 0.75 and a post-eclipse hump centred at 
phase $\sim$0.2. It also displays a flat region with differential R-magnitude 
$\sim$0.75, which extends from phase $\sim$0.25 to $\sim$0.5. We will consider
this flat region as the out-of-eclipse light level. In Fig.~\ref{fig3} 
we show a triple-Gaussian fit to the average light curve, assuming that the 
light curve can be decomposed into four components: a constant out-of-eclipse 
intensity, a hump, a wide dip and the eclipse. We want to stress that this triple-Gaussian fit
is merely an aid to the eye, just included to clarify the light curve structure. The presence of the wide dip in 
brightness prior to eclipse added to the strong flickering present in the light 
curves may have prevented Hellier et al. (\cite{hellier94b}) from detecting the 
eclipse in one of their V-band light curves.

The intricate shape of the light curve makes an explanation of the nature of its components very difficult, but we
will suggest some possibilities. Due to the lack of adequate spectroscopic observations (i.e. the spectral
features of the secondary
star are not detected in the optical), it could not be clear whether the eclipse center corresponds to inferior
conjunction of the secondary star. The possibility of the eclipse being a partial eclipse of the bright spot can
not be {\em a priori} ruled out, but there are evidences against it: (i) the eclipse is not flat-bottomed, as seen
in e.g. \object{U Gem} (Zhang \& Robinson \cite{zhang87}), although the observed V-shaped profile can arise from a
grazing eclipse of the bright spot; (ii) if our zero phase corresponds to an eclipse of the bright spot, the
brightening at $\varphi \sim$0.9 should correspond to a pre-eclipse orbital hump. This requires an out-of-eclipse
level located at the bottom of the dip at $\varphi \sim$0.75. This intensity level should be observed during a more
extended range of phases than the mean light curve shows (i.e. the bottom of the dip is narrow); (iii) the
eclipse of the bright spot should occur after the pre-eclipse orbital hump, when the intensity is decreasing
towards the out-of-eclipse level (see again Zhang \& Robinson \cite{zhang87}) . In \object{WX Ari} the eclipse is observed when the brightness is increasing.
In the light of these facts, we consider that the eclipse is more likely a partial eclipse of the disc, and is
centred at inferior conjunction of the secondary.

The light curve structure resembles that of other eclipsing CVs in the optical domain, like the nova-likes \object{UX UMa} and
\object{RW Tri} (Mason et al. \cite{mason97}) and is also similar to the X-ray light curves of some LMXBs, like
\object{X1822-371} (Hellier \& Mason \cite{hellier89}).
This similarities lead us to propose that the wide dip in brightness observed in \object{WX Ari} before eclipse may be caused by a
raised disc rim downstream from the bright spot (Armitage \& Livio \cite{armitage98}). When it is on the line of 
sight between the disc centre and the observer, it obscures a significant 
fraction of the disc causing the observed dip. A similar dip was observed in the dwarf 
nova \object{Z Cha} in the phase range 0.6--0.8 during superoutburst (Kuulkers 
et al. \cite{kuulkers91}) as well as in other high-inclination dwarf novae (e.g. 
\object{OY Car}, Naylor et al. \cite{naylor87},
; \object{U Gem}, Mason et al. \cite{mason88}) and other
CVs like \object{EX Hya} (C\'ordova et al. \cite{cordova85}), \object{BG CMi} (McHardy et al. 
\cite{mchardy86}). The observed dip extends nearly half of the
orbital cycle, which implies that the structure causing it may extend almost halfway around the disc. The same extention is seen in \object{RW Tri} (Mason et al. \cite{mason97}).

The nature of the post-eclipse hump is the most puzzling matter. It may be due 
to the fact that at phase $\sim$0.2 we are viewing the inner side of the raised 
rim, which may be reprocessing energetic radiation coming from the inner parts 
of the disc into visible wavelengths. The problem is that the widths of the dip and the hump are very different
and there is no strong reason to expect the reprocessed radiation to originate only in a small region of the
inside extended rim. The small width of this feature suggests that it may be related to the emergence of the bright spot out of
the secondary, but with the current data we can conclude nothing with respect to its origin.

\subsection{Orbital inclination}

From the geometry of a point eclipse by a spherical body, it is possible to 
determine the inclination, $i$, of a binary system through the relation

\begin{equation}
\label{eq2}
\left({R_2} \over 
{a}\right)^2=\sin^2(\pi\Delta\varphi_{1/2})+\cos^2(\pi\Delta\varphi_{1/2})\cos^2 i,
\end{equation}

\noindent
where $R_2/a$ is the volume radius of the secondary star, which depends only on 
the mass ratio, $q=M_2/M_1$ (Eggleton \cite{eggleton83}):

\begin{equation}
\label{eq3}
{{R_2} \over {a}}={{0.49~q^{2/3}} \over {0.6~q^{2/3}+\ln(1+q^{1/3})}}.
\end{equation}

$\Delta\varphi_{1/2}$ is the mean phase full-width of the eclipse at half the out-of-eclipse
intensity. In order to determine the orbital inclination of \object{WX Ari}, we calculated $\Delta\varphi_{1/2}$
from two different Gaussian fits: (i) the triple-Gaussian
fit to the mean light curve described above and (ii) a single-Gaussian fit to the eclipse profile
after masking the rest of the light curve, with the exception of the flat region extending from phase
0.25 to 0.5. The adopted $\Delta\varphi_{1/2}$ was the average value of those
derived from the two different fits, which is 
$$\Delta\varphi_{1/2}=0.082\pm0.006$$ 

Eliminating $R_2/a$ from Eqs. (\ref{eq2}) 
and (\ref{eq3}), we can obtain an estimate of the orbital inclination of the system assuming a
value of the mass ratio, $q$. Using the mass-period relation recently derived by Smith \& Dhillon 
(\cite{smith98}):

\begin{equation}
\label{eq4}
M_2(M_\odot)=0.126~P_{orb}({\rm h})-0.11,
\end{equation}

\noindent
where $P_{orb}$(h) is the orbital period of the system expressed in hours, we 
obtain a value of $M_2=0.31~M_\odot$ for the secondary in \object{WX Ari}. Assuming a mass for the
compact object of $M_1=0.8~M_\odot$ (the average mass of white dwarfs in CVs with orbital periods
above the period gap;  Smith \& Dhillon \cite{smith98}), we derive a value of the mass 
ratio for \object{WX Ari} of $q=0.39$. This mass ratio corresponds to an orbital 
inclination of $i\simeq80^{\rm o}$. Taking into account the error in $\Delta\varphi_{1/2}$ the
orbital inclination ranges between $\sim79^{\rm o}$ and $\sim82^{\rm o}$, {\em if the
eclipsed light source is centred on the white dwarf}.

In (\ref{eq2}) the parameter $a$ is the distance from the eclipsed light source to the center
of the secondary star, whereas in (\ref{eq3}) it is the separation of the
system. The shallowness of the eclipse in \object{WX Ari} suggests that we are actually observing a
partial eclipse of the disc (i.e. the white dwarf is not eclipsed), so $a$ is not the same quantity
in both Eqs. and we are in fact overestimating the inclination. Hence, the value 
derived above ($i\simeq80^{\rm o}$) can be considered as a (very pesimistic) upper limit to the
orbital inclination.

\subsubsection{The brightness of the discs in \object{SW Sex} stars}

The absolute magnitude of an steady state accretion disc is a function of
many variables (Warner \cite{warner87}), $M_V=M_V(\dot{M}_d, M_1, R_1, R_d)$, where
$\dot{M}_d$
is the mass accretion rate through the disc, $M_1$ is the mass of the white dwarf, $R_1$
its radius and $R_d$ is the radius of the disc. By adopting an $M-R$ relation for the
primaries in CVs, the dependence on $R_1$ can be eliminated. We assume the mass of the white dwarfs in
\object{SW Sex} systems,
$M_1$, as a constant. On the other hand, the radius of the discs in these systems does not differ very
much from one object to the others (Harrop-Allin \& Warner \cite{harrop96}) and can be taken
as $R_d/a=0.6-0.7~ R_{L_1}/a$, where $R_{L_1}/a$ is the distance (in units of the separation of the system)
from the centre of the primary to
the inner Lagrangian point. $R_{L_1}/a$ only depends on the mass ratio, $q$
(see e.g. Warner \cite{warner95}). From Kepler's Third Law
we obtain that $R_d=R_d(M_1, P_{orb})$ and then we have $M_V=M_V(\dot{M}_d, P_{orb})$. 
The eclipsing \object{SW Sex} systems are confined in a short orbital period range of 3.24--3.95
hours. The result of entering this short period range in $\dot{M}-P_{orb}$ relations (e.g.
Verbunt \& Wade \cite{verbunt84}) is a nearly equal $\dot{M}_d$ in the eclipsing \object{SW
Sex} stars, so we can assume that $M_V$ is only function of $P_{orb}$. Thus, the $M_V$ of the
discs in \object{SW Sex} stars must be approximately the same. As a consequence, the eclipse depth must be
a function of the orbital inclination.

\begin{figure}[!tbp]
  \resizebox{\hsize}{!}{\includegraphics{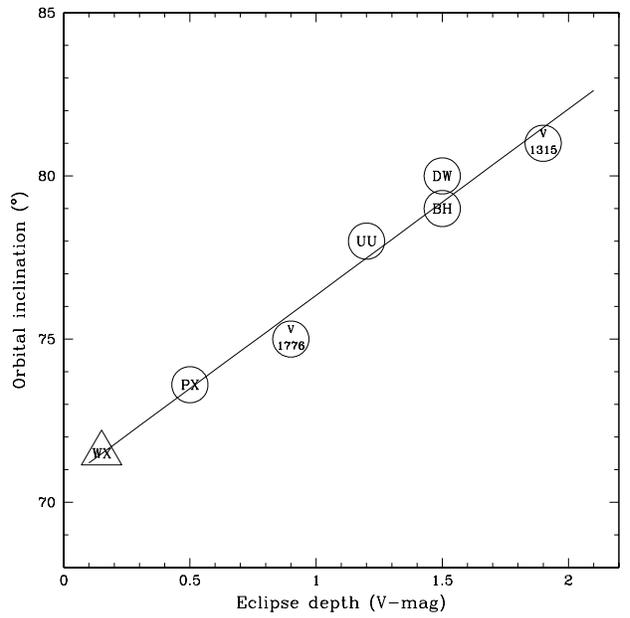}}
   \caption{Orbital inclination as a function of V-band eclipse depth for
   \object{SW Sex} stars. \object{WX Ari} is plotted with a triangle because the eclipse depth was measured
   in the R-band (see text for details).}
   \label{fig4}
\end{figure}

We have searched for values of orbital inclination and eclipse depth for all the eclipsing \object{SW
Sex} stars in the literature, with the aim of finding a correlation. In Fig.
\ref{fig4} we present a plot of the values found, which are summarized in Table~
\ref{table3}.

   \begin{table}[!tbp]
      \caption[]{V-band eclipse depth and $\Delta\varphi_{1/2}$ of \object{SW Sex}
      systems}
         \label{table3}
      \[
         \begin{array}{lccc}
            \hline
            \noalign{\smallskip}
            \rm Object      &  \Delta V  &  {\rm i}~(^{\rm o})  & \rm Reference \\
            \noalign{\smallskip}
            \hline
            \noalign{\smallskip}
            \rm PX~And     & 0.5 & 73.6 & 1     \\
            \rm V1776~Cyg  & 0.9 & 75   & 2     \\
	    \rm UU~Aqr     & 1.2 & 78   & 3     \\
	    \rm DW~UMa     & 1.5 & 80   & 4     \\
	    \rm BH~Lyn     & 1.5 & 79   & 5     \\
	    \rm V1315~Aql  & 1.9 & 81   & 6     \\
            
            \noalign{\smallskip}
            \hline
         \end{array}
      \]
{\it References}: 1 Thorstensen et al. \cite{thorstensen}; 2 Garnavich et al.
\cite{garnavich90}; 3 Baptista et al. \cite{baptista94}; 4 Shafter et al. \cite{shafter88}; 5
Hoard \& Szkody \cite{hoard97}; 6 Dhillon et al. \cite{dhillon91}

   \end{table}

The eclipse depth for all the systems was measured from V-band light curves. The systems
\object{SW Sex} and \object{LX Ser} are not included in the plot due to the lack of
reasonable estimates of the orbital inclination. We found a clear linear correlation between
orbital inclination and eclipse depth. After fitting a straight line to the data we get:

\begin{equation}
\label{eq5}
i=5.7~\Delta V+70.6
\end{equation}

The fact that all the systems lie very well within a straight line, indicates that the statements derived
above are valid, particularly, $M_1$ and $M_V$ can be taken as constants. We do not have extensive V-band
photometry to measure the eclipse depth, but due to the low inclination of \object{WX Ari} we can consider
the eclipse depth in R-band not to be very different from that in V-band. Entering the eclipse depth
measured in \object{WX Ari} ($\sim0.15$ mag) in
(\ref{eq5}), we obtain a value of the orbital inclination of $i\sim72^{\rm o}$. We have to
note that the ordinate at the origin of the linear fit is 70 degrees, value below which
eclipses are not expected to occur.

\begin{acknowledgements}
      We are specially grateful to the Observatorio del Teide Telescope Manager, 
Dr. Francisco Garz\'on L\'opez, and to Dr. Oriol Fuentes Masip for allocating 
the observing time at the IAC 80 telescope and the Optical Ground Station (OGS), 
respectively. We thank Cristina Zurita, Javier Licandro and David Alcalde for performing some
of the observations. We also wish to thank the Observatory's support astronomers
Dr. Mar\'\i a Rosa Zapatero Osorio and Dr. Alejandro Oscoz, and the night assistants 
Ricard Casas, Luis Chinarro, \'Angel G\'omez, Santiago L\'opez and Luis Manad\'e
for their valuable help at the telescopes. We thank the referee for his/her helpful comments, which have
improved the clarity and quality of this work.
\end{acknowledgements}

\end{document}